# Contribution of photogenerated charge carriers to photothermal effect in optically opaque semiconductor samples

Milica Dragas [1,2,*], Slobodanka Galovic [3], Katarina Djordjevic [3,*]

1 Faculty of Philosophy, University of East Sarajevo, Vuka Karadžica 30,
71126 Lukavica, Republic of Srpska, Bosnia and Herzegovina; milica.dragas@ff.ues.rs.ba
2 Faculty of Physics, University of Belgrade, Studentski trg 12, 11001 Belgrade, Serbia
3 Vinca Institute of Nuclear Sciences—National Institute of the Republic of Serbia, University of Belgrade,Mike Petrovica Alasa 12-14, P.O. Box 522, 11001 Belgrade, Serbia;

* Correspondence: milica.dragas@ff.ues.rs.ba
katarina.djordjevic@vin.bg.ac.rs

## Abstract

As a consequence of laser radiation absorption, the photothermal response of semiconductors is governed by both the thermalization of the crystal lattice and the dynamics of photogenerated charge carriers. This carrier-related contribution leaves a characteristic signature in the temperature field through carrier diffusion and recombination, making the photothermal response sensitive to electronic transport and recombination properties. This sensitivity provides the basis for using photothermal methods to characterize the electronic properties of semiconductors. In this paper, we develop a theoretical model of the photothermal response of moderately doped opaque semiconductors under harmonically modulated laser excitation, accounting for the generation, diffusion, and recombination of minority charge carriers. We analyze how carrier lifetime, diffusion length, and surface recombination velocity affect the resulting temperature field and, consequently, the photothermal signal. In particular, we show that the thermal contribution associated with carrier recombination remains nonzero even when the surface recombination rate vanishes and for short carrier lifetimes, demonstrating that the signature of photogenerated carriers may persist under conditions where their contribution might otherwise be expected to become negligible. The results provide guidance for both the design and interpretation of photothermal and photoacoustic experiments, including the choice of modulation frequency and detection geometry, to enhance the sensitivity to carrier-related thermal signatures and improve the determination of electronic properties of semiconductor materials and devices.

## Introduction

Photothermal and photoacoustic techniques have become important non-destructive tools for the characterization of semiconductor materials and devices because they provide access to thermal, optical, and electronic properties through the analysis of the response to periodically modulated optical excitation [1-9]. In addition to determining thermophysical parameters, these methods have been increasingly employed for the evaluation of electronic transport properties, such as minority-carrier lifetime, diffusion length, and surface recombination velocity [10-16], which are of central importance for the design and optimization of optoelectronic, photonic, and nanoelectronic devices [17-21].

When a semiconductor is illuminated by modulated laser radiation with photon energy comparable to or larger than the band-gap energy, the absorbed optical energy is not converted into heat through a single mechanism [2,3,22-27]. One part of the absorbed energy is transferred almost instantaneously to the crystal lattice through carrier thermalization, while another part is temporarily stored in photogenerated charge carriers and released later through recombination processes [2,3]. Consequently, the measured photothermal signal represents the combined effect of direct thermalization and carrier-mediated heat generation [38-30].

Despite extensive studies, the relative contributions of instantaneous thermalization and delayed carrier recombination to the measured photothermal response have not been fully clarified, particularly in optically opaque semiconductors. In particular, it is not fully understood under which conditions the contribution of carrier recombination can be distinguished from the contribution associated with thermal diffusion, how this separation depends on the carrier lifetime and diffusion length, and which frequency ranges provide the highest sensitivity to electronic transport parameters. In addition, simplified models often neglect carrier-related contributions or assume that they disappear when the surface recombination velocity is zero. Such assumptions obscure the physical mechanisms responsible for the formation of the photothermal response and may limit the reliability of parameter extraction from experimental measurements. These unresolved issues become particularly significant in optically opaque semiconductors, where carrier transport and heat diffusion occur over comparable characteristic length scales.

In optically opaque semiconductors, where the optical absorption length is much smaller than the sample thickness, the interplay between carrier diffusion, recombination, and thermal diffusion becomes particularly important. The characteristic diffusion length of minority carriers determines whether photogenerated carriers can explore the entire sample volume before recombination or remain localized near the illuminated surface. As a consequence, the relative contributions of thermalization and recombination are expected to depend not only on the modulation frequency, but also on the relationship between the carrier diffusion length and the sample thickness. Unlike previous theoretical treatments, the present model explicitly separates heat generation due to carrier thermalization from that associated with bulk recombination. This formulation enables the individual contributions of the two mechanisms to be analyzed as functions of modulation frequency, carrier lifetime, diffusion length, and sample thickness.

The objective of this work is to investigate the contribution of photogenerated minority carriers to the photothermal response of optically opaque semiconductor samples and to identify the conditions under which carrier-related effects can be separated from purely thermal contributions. To this end, a theoretical model is developed that simultaneously accounts for direct thermalization and bulk recombination as distinct heat-generation mechanisms. The influence of carrier lifetime, diffusion length, sample thickness, modulation frequency, and observation side of the sample on the amplitude and phase characteristics of the photothermal response is systematically analyzed.

Special attention is devoted to identifying frequency ranges and experimental configurations that maximize sensitivity to electronic transport properties. The obtained results provide physical insight into the formation of photothermal signals and establish guidelines for the design of photothermal and photoacoustic experiments aimed at the non-destructive electrical characterization of semiconductor materials and devices.

## 1. Theoretical framework

The considered geometry is shown in Figure 1. An optically opaque semiconductor sample of thickness $l$, is illuminated by harmonically modulated laser radiation $I_{abs} = \frac{I_0}{2}\left(1+\cos\left(\omega t\right)\right)e^{-\beta x}$ incident on one surface of the sample $x=0$. Since the optical absorption coefficient $\beta$ is assumed to be sufficiently large $\beta \to \infty$, the optical absorption depth is much smaller than the sample thickness $l$. Therefore, the samples are considered optically opaque, and the Beer–Lambert absorption profile is approximated by $e^{-\beta x} \approx \delta\left(x\right)/\beta$, where $\delta(x)$ is the Dirac delta function. Consequently, the absorbed optical energy is confined to a thin region near the illuminated surface, where both the generation of minority charge carriers $\delta n_p$ and the initial conversion of optical energy into heat take place [2,3,9,11].

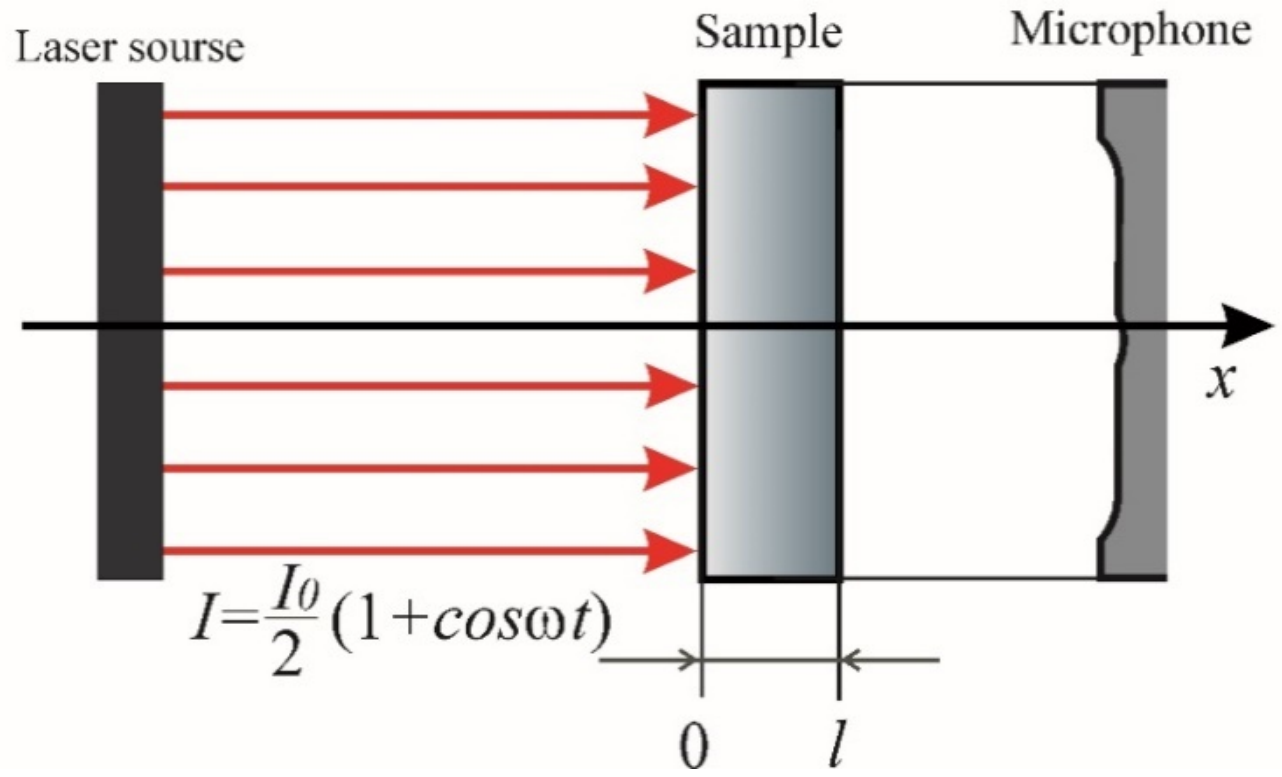

Figure 1. Geometry of the considered system

The characteristic times associated with optical absorption and carrier thermalization are assumed to be much shorter than the modulation period of the excitation source. Under these conditions, optical absorption may be considered instantaneous, and the carrier-generation rate follows the temporal modulation of the incident optical flux [31]. Likewise, the fraction of the absorbed optical energy transferred directly to the crystal lattice through thermalization is assumed to follow the harmonic excitation without delay [2,3]. This approximation is commonly employed in photothermal modeling of semiconductors and enables the optical excitation to be represented by two coupled source terms: a thermalization source acting directly on the lattice and a carrier-generation source governing the evolution of photogenerated minority carriers [2,3,10,32-36].

Because the optical excitation is spatially uniform over the illuminated surface and the sample thickness $l$ is much smaller than the lateral dimensions $l$<<$Rs$, heat and carrier transport are described within a one-dimensional approximation. The analysis therefore focuses on transport processes along the direction normal to the illuminated surface [32].

The photothermal response is described as a cascade of diffusion processes. First, photogenerated minority carriers $\delta n_p$ diffuse through the semiconductor volume under the action of a surface generation source $G=-\frac{1}{h\nu}\frac{\partial I_{abs}}{\partial x}$. Their spatiotemporal distribution $\delta n_p(x,t)$ gives rise to a volumetric recombination heat source $R=\frac{\delta n_p(x,t)}{\tau_p}$ whose intensity depends on the local minority carrier lifetime $\tau_p$, [2,3,33-38]. The resulting thermal field $\vartheta(x,t)$ determined by the combined contribution of heat-generation mechanisms: direct thermalization $\vartheta_T(x,t)$ localized near the illuminated surface, and bulk carrier recombination $\vartheta_{BR}(x,t)$ distributed throughout the sample volume and surface carrier recombination $\vartheta_{SR}(x,t)$ occurring at the surfaces:

$$\vartheta(x,t)=\vartheta_T(x,t)+\vartheta_{BR}(x,t)+\vartheta_{SR}(x,t) \tag{1}$$

In addition, the sample is surrounded by air, whose thermal conductivity is significantly lower than that of the semiconductor. Heat exchange with the surroundings is therefore neglected and adiabatic Neumann boundary conditions are adopted for the temperature field as well [38-44]. Under these assumptions, the measured photothermal response $\vartheta(x,t)$ is determined exclusively by the internal generation, diffusion, and recombination processes occurring within the semiconductor sample.

## 2.1 Governing equations and excess charge carrier distribution

The time-periodic evolution of the excitation $\overline{I}=\frac{I_0}{2}\left(1+e^{i\omega t}\right)$, characterized by the modulation frequency $\omega$ and incident beam intensity $I_0$, in the n-type Si semiconductor sample (Figure 1), is accompanied by absorption and energy conversion. Thermalization is converted into heat almost instantaneously, whereas the transformation of a fraction of the absorbed energy into photogenerated carriers (electrons and holes) is a spatiotemporally dependent process obtained as the solution of the diffusion equation describing the propagation of carriers generated by the surface, time-periodic excitation:

$$\left[\frac{\partial^2}{\partial x^2}-\frac{1}{D_p}\frac{\partial}{\partial t}\right]\delta n_p(x,t)=-\frac{G-R}{D_p} \tag{2}$$

with $D_p$ diffusivity of minority carriers, and $\delta n_p(x,t)$ the minority-carrier concentration is significant during the dynamics of processes in n-type silicon, while electrons cause only a negligible change in the majority-carrier concentration. The generated carriers $G$ diffuse throughout the sample. Their recombination $R$ releases heat in the bulk of the sample, while recombination at the sample surfaces releases heat at both the illuminated and unilluminated surfaces. Since the semiconductor is electrically isolated and no carrier flux crosses the sample boundaries [42], homogeneous Neumann boundary conditions [33,43] are imposed on the minority-carrier concentration:

$$D_p \left.\frac{\partial \delta n_p(x,t)}{\partial x}\right|_{x=0} = s_1 \delta n_p(0,t), \qquad D_p \left.\frac{\partial \delta n_p(x,t)}{\partial x}\right|_{x=l} = -s_2 \delta n_p(l,t) \tag{3}$$

With $s_1$, $s_2$ surface recombination rate on illuminated and unilluminated surface of the sample, respectively.

To solve the partial differential equation describing the excess charge carrier concentration, the excess carrier concentration and the intensity of the modulated light are represented by the complex phasors $\delta n_p(x,t) = \frac{\delta \overline{n}_p(x)(1+e^{i\omega t})}{2}$, $\overline{I} = \frac{I_0}{2}\left(1+e^{i\omega t}\right)$, respectively. Since the incident light intensity is modulated at the angular frequency ω, the photogenerated charge carriers are also modulated at the same angular frequency. By introducing the complex harmonic representation into the differential equation, the problem is separated into its static and dynamic components, yielding two differential equations. The equation of interest with a complex representation is a dynamic equation; homogeneous differential equations in the complex domain:

$$\left[\frac{d^2}{dx^2} - \frac{1}{L^2}\right] \delta \overline{n}_p(x) = 0 \tag{4}$$

With inhomogeneous boundary conditions that incorporate recombination at the illuminated surface $x = 0$, and at the unilluminated surface of the sample $x = l$:

$$D_p \left.\frac{d \delta \overline{n}_p(x)}{dx}\right|_{x=0} = s_1 \delta \overline{n}_p(0) - \frac{I_0}{h\nu}, \qquad D_p \left.\frac{d \delta \overline{n}_p(x)}{dx}\right|_{x=l} = -s_2 \delta \overline{n}_p(l) \tag{5}$$

where $L = \sqrt{\frac{D_p \tau_p}{1 + i\omega\tau_p}}$ complex diffusion length of the minority carriers.

Solution of eg. (4) with boundary condition is given in form of linear combination:

$$\delta \overline{n}_p(x) = A_1 e^{x/L} + A_2 e^{-x/L} \quad , \quad \text{where:} \tag{6}$$

$$
\begin{cases}
A_2 = \dfrac{c(1+a_2)e^{l/L}}{(1+a_1)(1+a_2)e^{l/L} - (1-a_1)(1-a_2)e^{-l/L}}, \\
A_1 = A_2 \dfrac{(1+a_1)}{(1-a_1)} - \dfrac{c}{(1-a_1)}, \\
a_1 = \dfrac{s_1 L}{D_p}, a_2 = \dfrac{s_2 L}{D_p}, c = \dfrac{I_0 L}{h\nu D_p}
\end{cases}
\tag{7}
$$

In this chapter, the spatial distribution of the minority carrier concentration in the sample is determined. The obtained analytical solution shows that the minority carrier concentration depends on the diffusion length and diffusivity of the minority carriers, and is expressed as a function of the depth and thickness of the sample. The analysis indicates that these parameters have a significant influence on the spatial distribution of minority carriers, providing a basis for further investigation of carrier transport processes and the analysis of the electrical characteristics of the considered structure.

2.2 Governing equation and photothermal response

The previous analysis determined the spatial and temporal distribution of photogenerated excess charge carriers under periodically modulated optical excitation. This distribution provides the basis for describing heat generation associated with carrier diffusion and recombination.

Photon absorption and carrier thermalization occur much faster than carrier diffusion and recombination. Therefore, the energy transferred to the lattice during thermalization can be treated as an instantaneous heat source, while the remaining absorbed energy is redistributed in space and time as the carriers diffuse and recombine. The carrier concentration n(x,t) is thus spatially and temporally dependent and is determined by the diffusion equation. Accordingly, the optical excitation can be described by two components: a spatially uniform component representing the instantaneous heat generation, and a spatially dependent, phase-shifted component describing the delayed heat generation due to carrier diffusion and recombination.

The thermodiffusion equation that describes temperature fild under the assumptions is obtained from a one-dimensional diffusion impulse partial differential equation with three characteristic contributions and zero boundary conditions.

$$
\left[\frac{\partial^2}{\partial x^2} - \frac{1}{D_T}\frac{\partial}{\partial t}\right]\vartheta_i(x,t) = -\frac{I_0(1+e^{i\omega t})}{k}\frac{h\nu - \varepsilon_g}{h\nu}\delta(x) - \frac{\varepsilon_g \delta n_p(x,t)}{k\tau_p} - \frac{s_1 \varepsilon_g \delta n_p(0,f)}{k}\delta(x) + \frac{s_2 \varepsilon_g \delta n_p(l,f)}{k}\delta(x-l)
$$

(8)

$$
-k\frac{\partial \vartheta_i(x,t)}{\partial x}\bigg|_{x=0} = 0, \qquad -k\frac{\partial \vartheta_i(x,t)}{\partial x}\bigg|_{x=l} = 0 \ , \tag{9}
$$

where $k$ denotes the thermal conductivity, $\varepsilon_g$ is the semiconductor bandgap energy, $h\nu$ is the energy of the incident photons, $D_T$ -thermal diffusivity.
The temperature distribution is obtained in an analogous manner as for the excess carrier concentration. The temperature field is represented by its complex harmonic form $\vartheta(x,t) = \frac{\bar{\vartheta}(x)(1+e^{i\omega t})}{2}$, which enables the governing equation to be decomposed into static and dynamic components. The resulting temperature field is evaluated by considering the contribution of each heat-generation mechanism separately.
The temperature distributions associated with thermalization, bulk recombination, and surface recombination are obtained by solving the following governing equations.

The temperature distribution due to the contribution of the thermalization:

$$\left[\frac{d^2}{dx^2} - \sigma^2\right]\bar{\vartheta}_T(x) = 0 \tag{1}$$

$$-k\frac{d\vartheta_T(x)}{dx}\bigg|_{x=0} = \frac{h\nu - \varepsilon_g}{h\nu} I_0, \qquad -k\frac{d\vartheta_T(x)}{dx}\bigg|_{x=l} = 0 \tag{11}$$

bulk recombination:

$$\left[\frac{d^2}{dx^2} - \sigma^2\right]\bar{\vartheta}_{BR}(x) = -\frac{\varepsilon_g \delta\bar{n}_p(x)}{k\tau_p} \tag{2}$$

$$-k\frac{d\vartheta_{BR}(x)}{dx}\bigg|_{x=0} = 0, \qquad -k\frac{d\vartheta_{BR}(x)}{dx}\bigg|_{x=l} = 0 \tag{3}$$

surface recombination:

$$\left[\frac{d^2}{dx^2} - \sigma^2\right]\bar{\vartheta}_{SR}(x) = 0 \tag{4}$$

$$-k\frac{d\vartheta_{SR}(x)}{dx}\bigg|_{x=0} = s_1\varepsilon_g\delta\bar{n}_p(0), \qquad -k\frac{d\vartheta_{SR}(x)}{dx}\bigg|_{x=l} = -s_2\varepsilon_g\delta\bar{n}_p(0) \tag{5}$$

With $\sigma = \sqrt{\frac{i\omega}{D_T}}$ - complex heat diffusion coefficient.

The temperature fild along the sample, eq (7), resulting from thermalization as well as bulk and surface recombination in optically opaque semiconductors, is determined by the following expressions for termalization:

$$\vartheta_T(x) = B_1 e^{\sigma x} + B_2 e^{-\sigma x}$$

$$\begin{cases} B_2 = \dfrac{h\nu - \varepsilon_g}{k\sigma h\nu} \dfrac{I_0 e^{\sigma l}}{e^{\sigma l} - e^{-\sigma l}} \\ B_1 = B_2 + \dfrac{h\nu - \varepsilon_g}{k\sigma h\nu} I_0 \end{cases} \tag{6}$$

For bulk recombindation:

$$\vartheta_{BR}(x) = C_1 e^{\sigma x} + C_2 e^{-\sigma x} + C_3 e^{x/L} + C_4 e^{-x/L}$$

$$\begin{cases} C_2 = \dfrac{C_3(e^{\sigma l} - e^{l/L}) - C_4(e^{\sigma l} - e^{-l/L})}{L\sigma(e^{\sigma l} - e^{-\sigma l})} \\ C_1 = C_2 - \dfrac{C_3 - C_4}{L\sigma} \\ C_3 = \dfrac{\varepsilon_g}{k\tau_p} \dfrac{A_1}{\sigma^2 - L^{-2}} \\ C_4 = \dfrac{\varepsilon_g}{k\tau_p} \dfrac{A_2}{\sigma^2 - L^{-2}} \end{cases} \tag{7}$$

And for surface recombination:

$$\vartheta_{SR}(x) = D_1 e^{\sigma x} + D_2 e^{-\sigma x}$$

$$\begin{cases} D_2 = \dfrac{\varepsilon_g}{k\sigma} \dfrac{s_1 \delta \overline{n}_p(0) e^{\sigma l} + s_2 \delta \overline{n}_p(l)}{(e^{\sigma l} - e^{-\sigma l})} \\ D_1 = D_2 - \dfrac{\varepsilon_g}{k\sigma} \delta \overline{n}_p(0) \end{cases} \tag{8}$$

In this chapter, a theoretical model is developed to describe the photothermal response in an optically opaque semiconductor under modulated optical excitation. Based on the proposed mathematical model, analytical solutions have been obtained for the thermalization, bulk recombination, and surface recombination contributions. These solutions enable an analysis of the influence of material parameters and excitation conditions on the individual contributions to the photothermal response in optically opaque semiconductors, which will be examined in the following chapters. The model also indicates that experimental conditions should be chosen to minimize the thermalization contribution and emphasize carrier recombination, thereby providing more reliable information on the electronic properties of the material.

## 3. Results and Analysis

This chapter presents the photothermal response of opaque semiconductor samples obtained using the analytical solutions derived in the previous chapter. The effects of thermalization and recombination processes on the sample temperature variation are analyzed.

Samples with ideally polished surfaces were considered, for which $s_1 = s_2 = 0$, thereby completely eliminating surface recombination and the associated photothermal response from the analysis. The results are presented as amplitude and phase characteristics of the photothermal response $\vartheta(x,t)$ (solid lines) over the frequency range from 10 Hz to 100 kHz, with the contributions of thermalization $\vartheta_T(x)$ (dot) and bulk recombination (dash) shown separately. By comparing the results for sample thicknesses and carrier lifetimes $\tau$ for $10^{-4}\,\mathrm{s}$ and $10^{-6}\,\mathrm{s}$, the influence of these parameters on the response characteristics, the dominant heat-generation mechanisms, and their relative contributions over the frequency range from 10 Hz to $10^7\,\mathrm{Hz}$ was investigated. The obtained results make it possible to identify the regimes in which individual mechanisms dominate, as well as the extent of their variation with sample thickness and carrier lifetime. This provides a basis for selecting optimal experimental conditions and for a more reliable interpretation of photothermal measurements in opaque semiconductor materials.

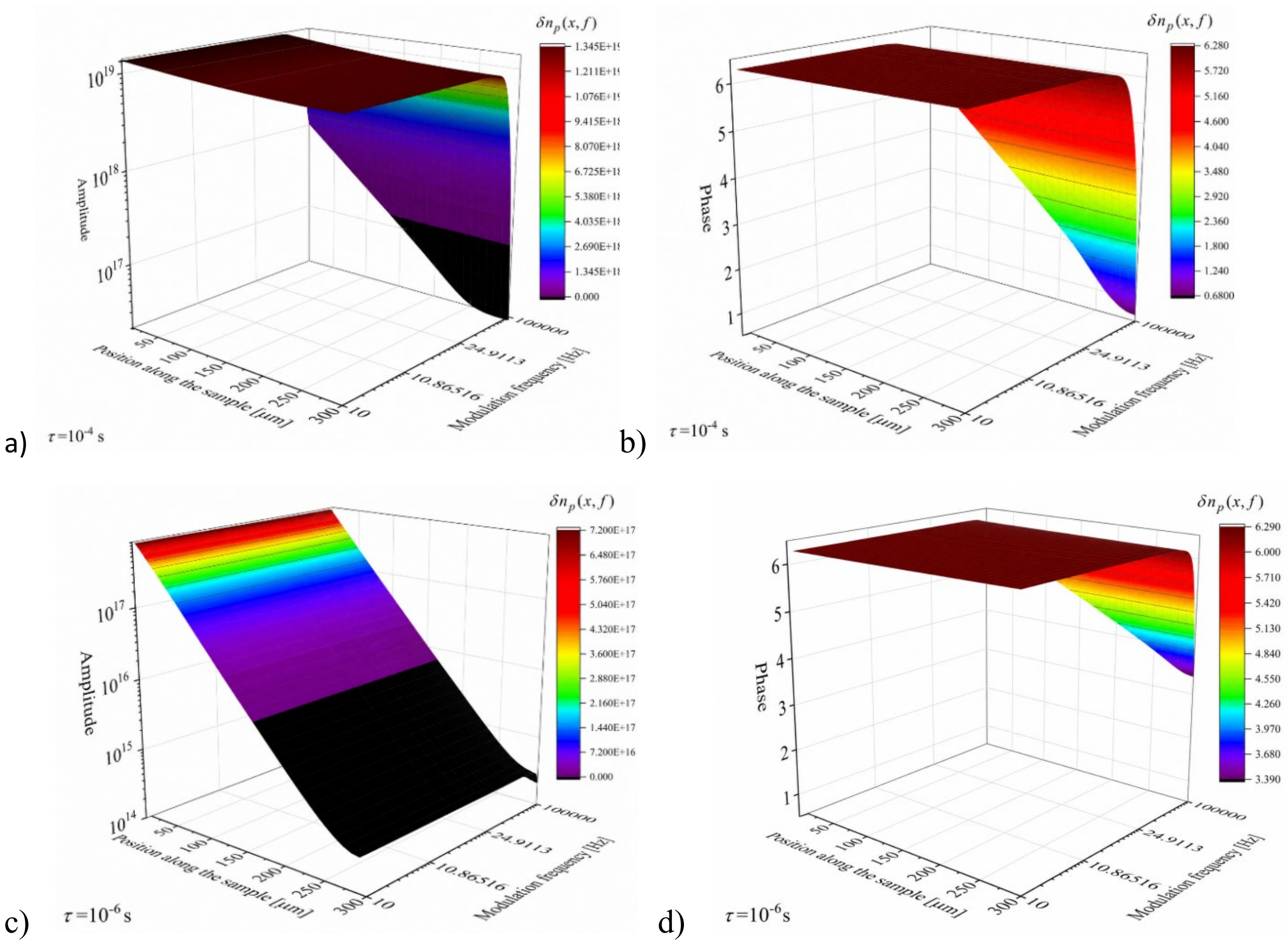

Figure 2 The concentration of photogenerated carriers given by Eqs. (6) and (7) is presented for a sample thickness of 300 μm at two relaxation times: in (a,b) for $\tau_1 = 10^{-4}\,\mathrm{s}$ and in (c,d) for $\tau_2 = 10^{-6}\,\mathrm{s}$.

The analysis was carried out for samples with thicknesses of 10 (black), 20 (red), 50 (green), 100 (blue), 200 (light blue), and 300 (pink) μm in order to examine the temperature field distribution and the influence of individual heat generation mechanisms from the change in sample thickness.. In addition, two characteristic carrier lifetimes, $\tau_1 = 10^{-4}$s and $\tau_2 = 10^{-6}$s, were considered. These values correspond to different bulk recombination conditions and enable an investigation of their influence on the temperature variation.

For the semiconductor with a carrier lifetime of $\tau_1 = 10^{-4}$s, the carrier diffusion length is $L_1$=346.41 μm. Since the thicknesses of all analyzed samples (10, 20, 50, 100, 200, and 300 μm) are smaller than the diffusion length $l < L$, all of the investigated samples are classified as plasma-thin. For the semiconductor with a carrier lifetime of $\tau_2 = 10^{-6}$s, the carrier diffusion length is $L_2$=34.64 μm. Under these conditions, the 10 and 20 μm samples satisfy $l < L$ and are therefore classified as plasma-thin samples, whereas the 50, 100, 200, and 300 μm samples, for which $l > L$, are classified as plasma-thick. The concentration of photogenerated carriers given by Eqs. (6) and (7) is presented for a sample thickness of 300 μm at two relaxation times: in Fig. 9(a,b) for $\tau_1 = 10^{-4}$s and in Fig. 9(c,d) for $\tau_2 = 10^{-6}$s

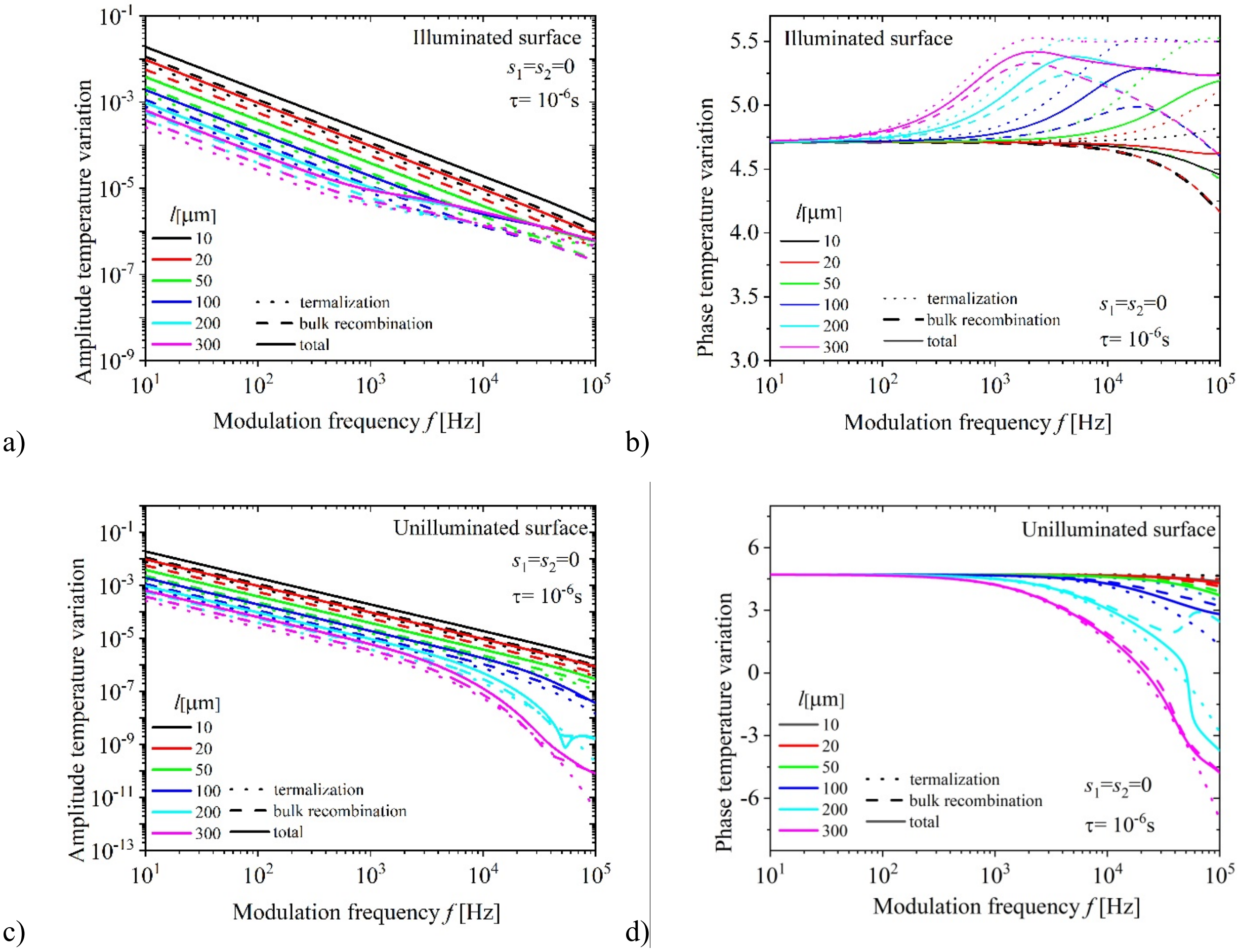

Figure 3. Thermalization (dot), bulk recombination (dash), and temperature variation (solid line) of silicon samples for a carrier lifetime of $10^{-6}$s: (a,c) amplitudes and (b,d) phases on the illuminated side of the sample (a,b) and on the non-illuminated side of the sample (c,d).

Amplitude and phase characteristics of the temperature field of the illuminated and non-illuminated surface of the opaque sample during the lifetime $\tau_2 = 10^{-6}$s were analyzed, Figure 3. Photothermal response of the illuminated side of the sample, has the highest values due to the influence of photogenerated charge carriers in the amplitude for plasma thin samples 10 and 20 in the entire frequency range, while in the amplitude of plasma thick samples at low frequencies from around 1kHz to 30kHz. The influence of thermalization in phases is most pronounced at frequencies >50Hz of plasma thick samples. The temperature variation of the non-illuminated surface of the sample has a significant influence of minority carriers up to frequencies of ~4kHz due to the dominance of bulk plasma recombination of thin samples. Attenuation of the influence of plasma carriers of thick samples leads to the smallest values of amplitudes and phases.

For plasma-thin samples, the contribution of bulk recombination is pronounced across the entire frequency range, whereas for plasma-thick samples, the contribution of thermalization on the illuminated side can be more clearly identified at higher frequencies, particularly through the phase characteristics, where the influence of charge carriers is less pronounced.

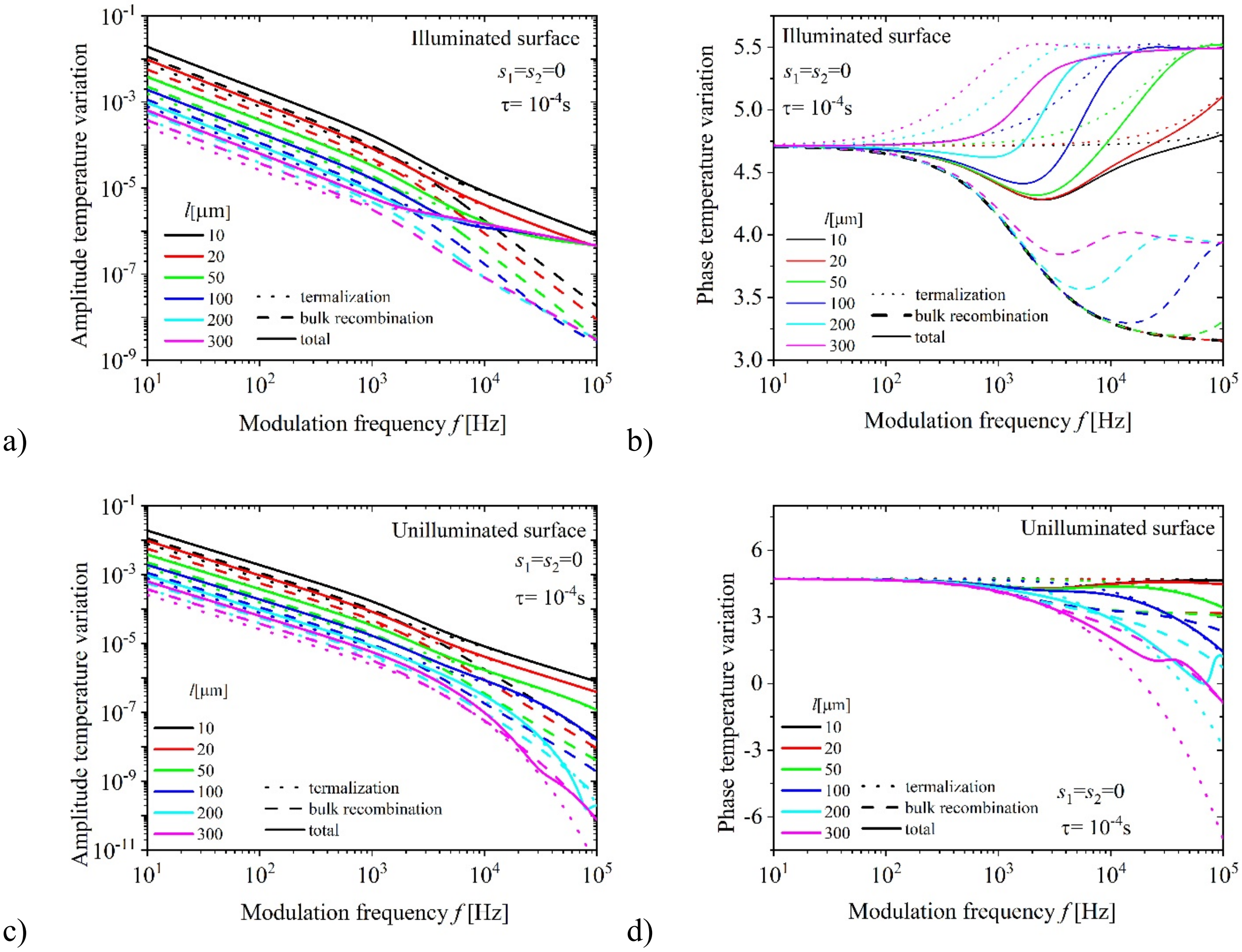

Figure 4. Thermalization (dot), bulk recombination (dash), and temperature variation (solid line) of silicon samples for a carrier lifetime of $\tau_1 = 10^{-4}$s : (a,c) amplitudes and (b,d) phases on the illuminated side of the sample (a,b) and on the non-illuminated side of the sample (c,d).

For a carrier lifetime of $\tau_1 = 10^{-4}\mathrm{s}$ , the amplitude characteristics of the temperature variation were analyzed on both the illuminated and non-illuminated sides of the sample, Figure 4a,c. In the low-frequency region, it is observed that the highest amplitude values are obtained for the thinnest samples, whereas the amplitude gradually decreases with increasing sample thickness, with the thickest samples exhibiting the lowest values. In the low-frequency range araund 1kHz to 3kHz, bulk recombination provides the dominant contribution to the temperature variation on both the illuminated and non-illuminated sides of the sample up to approximately 1kHz to 100 kHz. With increasing frequency, a change in the dominant mechanism occurs. In the high-frequency region, thermalization becomes the dominant contribution for almost all analyzed samples and signals. The exception is observed for the samples with thicknesses of 200 μm and 300 μm , for which bulk recombination becomes the dominant mechanism again over thermalization at frequencies above approximately 20 kHz.

Table 1. Frequency-dependent contributions of thermalization and bulk recombination for different carrier lifetimes and sample thicknesses

| τ | Low frequencies | High frequencies | Most pronounced carrier effect |
|---|---|---|---|
| $10^{-6}$ s | Recombination (thin samples) | Thermalization (thick, illuminated side) | Phase, illuminated side of thick samples |
| $10^{-4}$ s | Recombination (all samples) | Thermalization, except for 200–300 μm, where recombination again dominates on the non-illuminated side amplitudes and phases | The amplitude of the high-frequency response depends on sample thickness. The largest phase is for thin samples and decreases with increasing thickness. |

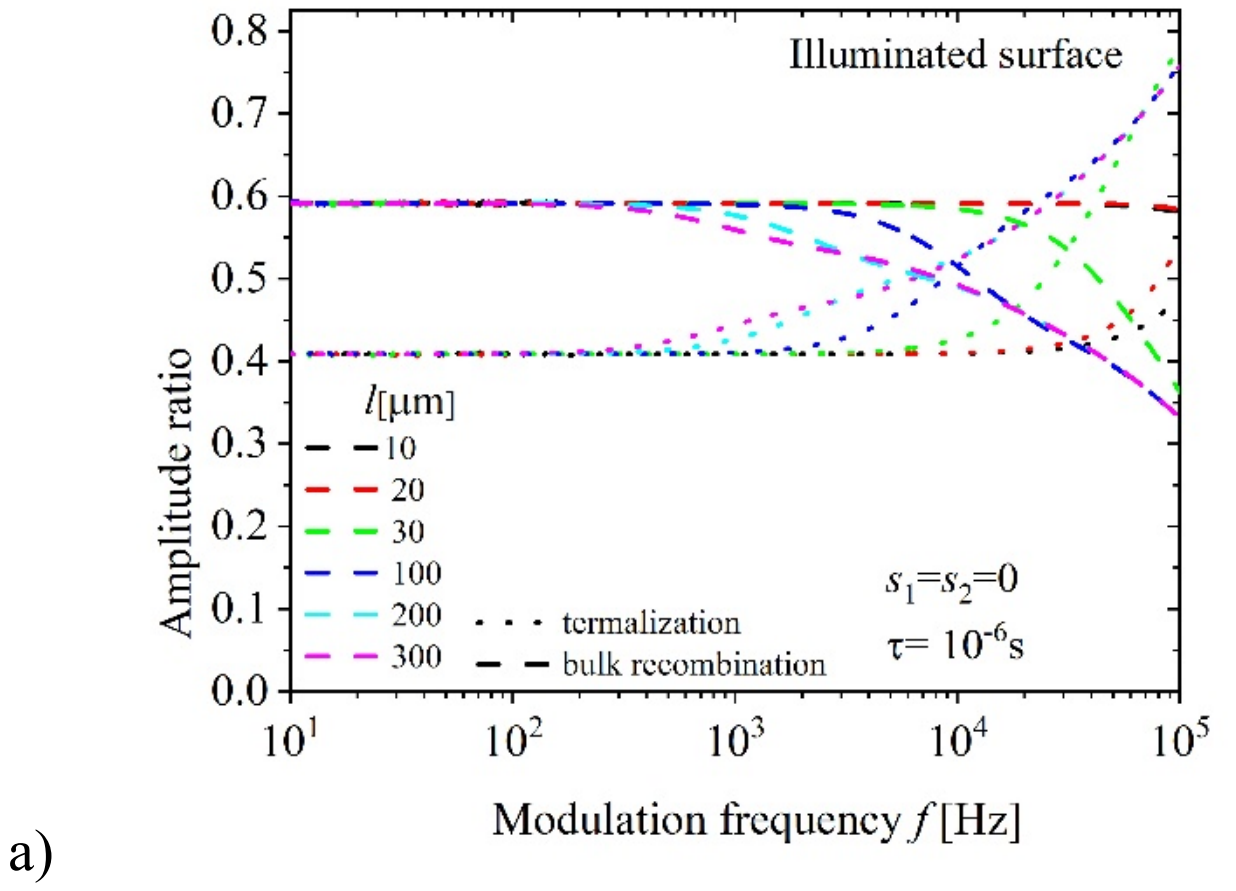

a)

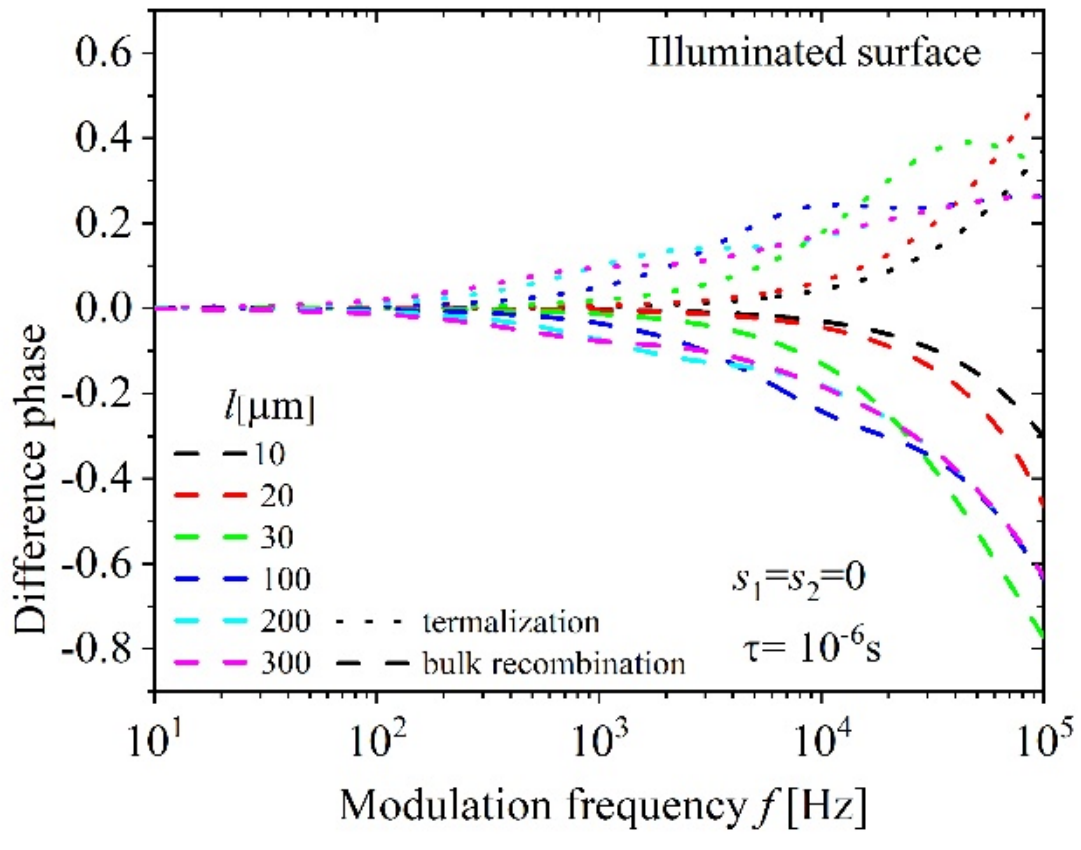

b)

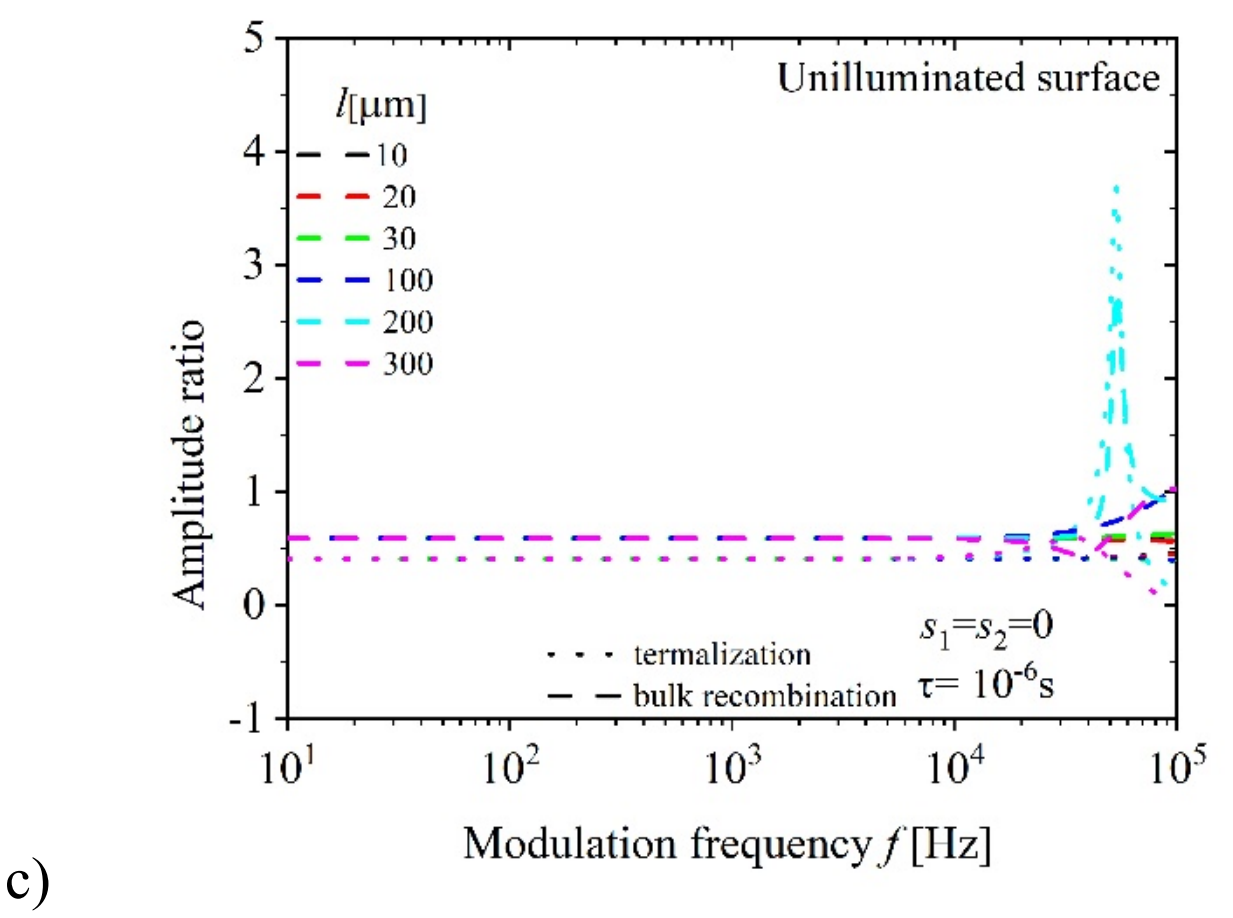

c)

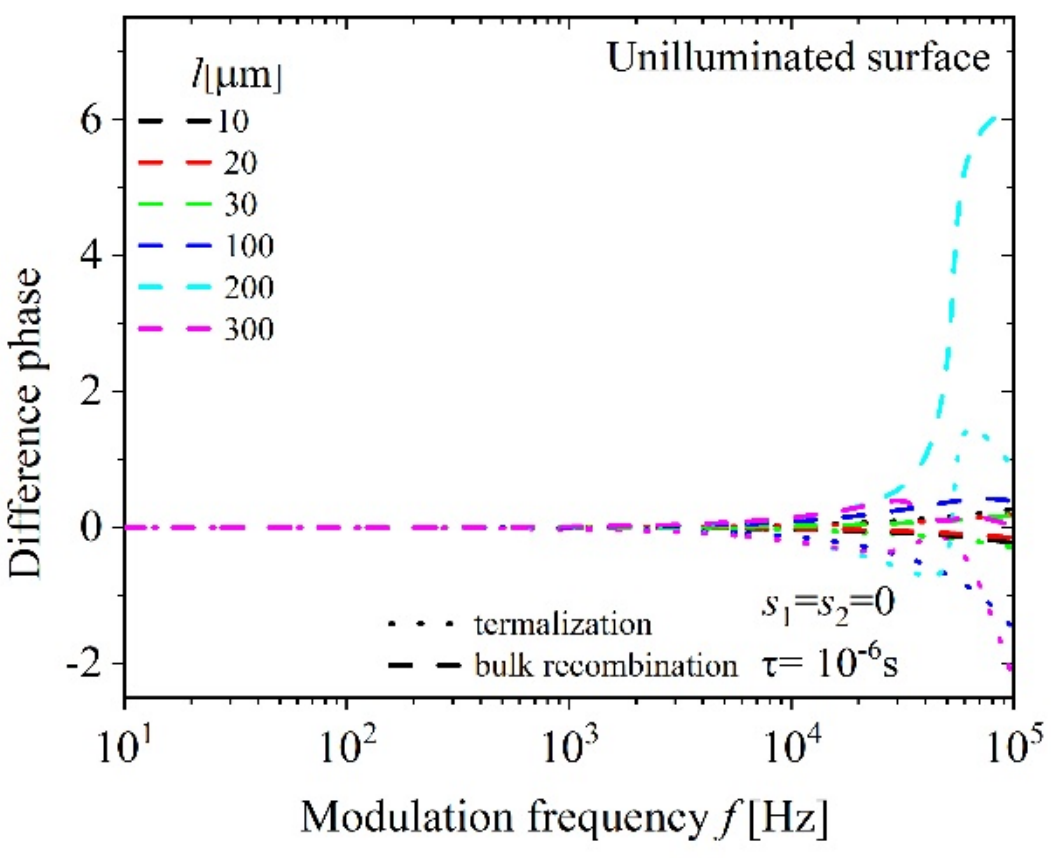

d)

Figure 5 Influence of thermalization (dot) and bulk recombination (dash) in the photothermal response of the sample during the life time of $\tau_2 = 10^{-6}\,\mathrm{s}$ : (a,c) amplitude ratio and (b,d) phase difference on the illuminated surface of the sample (a,b), and on the non-illuminated surface of the sample (c,d).

Figure 5 shows the contribution of thermalization and bulk recombination to the photothermal response with lifetime $\tau_2 = 10^{-6}\,\mathrm{s}$ . At low frequencies in the amplitudes of the illuminated surface, Figure 5a, recombination is dominant with ~60% compared to thermalization ~40% up to 300Hz. The change of recombination dominance in relation to thermalization is manifested in thicker samples in the medium frequency range up to 30kHz. In the amplitudes of the unilluminated side, Figure 5c, the dominance of recombination is maintained up to 30 kHz, after which an anomaly appears in thicker samples (appearance of a peak for the 200 $\mu\mathrm{m}$ sample, and recombination increase again for 300 $\mu\mathrm{m}$). In other samples, the changes are small. In the phases of the illuminated sample surface, Figure 5b, the region of equal contributions of thermalization and recombination up to ~100Hz, after which thermalization becomes dominant. The effect appears first in thicker samples. In the phases of the non-illuminated surface of the sample, Figure 5d, the area of equal contributions of thermalization and recombination is up to ~10 kHz, after which, in thicker samples, the contribution of recombination increases, and the contribution of thermalization decreases.

The analysis of the amplitude ratio and phase difference shows that volume recombination is the dominant mechanism in the low-frequency region, while with the increase in frequency, the dominance of thermalization is gradually established on the illuminated side of the sample, especially in thicker samples, Figure a, c. In contrast, on the unilluminated side, the contribution of volume recombination remains significant in a much wider frequency range, and in the case of the thickest samples, it is even stronger again in the high-frequency area, Figure 5c. This indicates that the illuminated side provides greater sensitivity to thermalization processes, Figure 5 a,b, while the unilluminated side is more suitable for identifying the volume recombination contribution, especially in thicker samples, Figure 5c,d

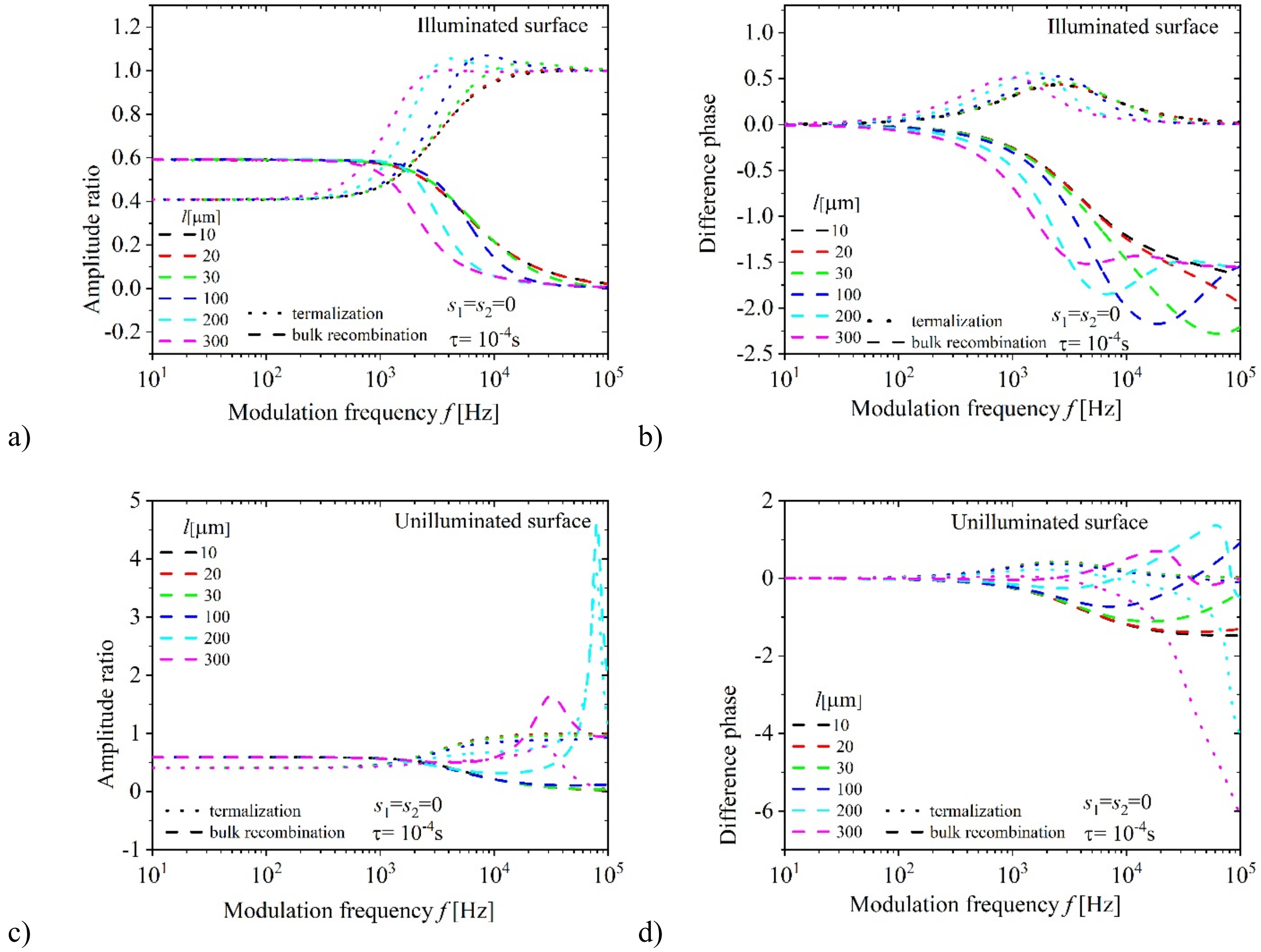

Figure 6. Contributions of thermalization (dot) and bulk recombination (dash) to the photothermal response for the opaque sample with carrier lifetime $\tau_1 = 10^{-4}\,\mathrm{s}$: (a,c) amplitude ratios and (b,d) phase differences on the illuminated surface of the sample (a,b) and on the non-illuminated surface of the sample (c,d).

Figure 6 shows the frequency contributions of thermalization and bulk recombination to the photothermal response for the carrier lifetime $\tau_1 = 10^{-4}\,\mathrm{s}$. At the amplitude of the illuminated surface of the sample (Figure 6a), bulk recombination is shown to be dominant over thermalization, with a ratio of 0.6 to 0.4, up to the frequency range of 800 Hz to 20 kHz, after which thermalization dominates in the high-frequency region. In the amplitude ratio of the non-illuminated side (Figure 6c), bulk recombination is shown to dominate up to 2 kHz, after which thermalization becomes dominant in thinner samples, whereas recombination peaks appear in thicker samples. Thicker samples enter the thermalization dominated regime at lower frequencies compared to thinner samples. A smooth transition from thermalization to recombination is observed. The most pronounced deviations occur for the 200 and 300 μm samples at high frequencies on the non-illuminated side, as shown in Figures 6a and c.

In terms of phase differences, the contributions of thermalization and bulk recombination are relatively equal up to approximately 50 Hz on the illuminated surface of the samples (Figure 6b), and up to approximately 200 Hz on the non-illuminated side of the sample (Figure 6d). Thermalization dominates at higher frequencies, from 200 Hz to 20 kHz on the illuminated surface and above 200 Hz on the non-illuminated surface of the sample. The most pronounced influence of bulk recombination on the illuminated side occurs from 2 to 100 kHz, after which a saddle-shaped profile develops, as shown in Figure 6b. In contrast, on the non-illuminated surface of the sample, the influence of bulk recombination at 100 and 200 Hz (Figure 6d) becomes apparent above 2 kHz, with non-monotonic recombination behavior observed in the thicker samples.

For a carrier lifetime of $\tau_1 = 10^{-4}\,\mathrm{s}$, both the amplitude ratio and phase difference exhibit a clear transition from bulk recombination dominance in the low-frequency region to thermalization dominance in the high-frequency region. This transition occurs at lower frequencies on the illuminated side than on the non-illuminated side of the sample. While the thinner samples exhibit an almost monotonic increase in the thermalization contribution, the thicker samples show more complex high-frequency behavior, with local maxima in the bulk recombination contribution, particularly on the non-illuminated side of the sample.

## 4. Discussion

The obtained results show that the photothermal response of moderately doped semiconductors results from the simultaneous action of two distinct mechanisms: direct thermalization of the crystal lattice and recombination of photogenerated minority charge carriers. The relative contribution of these two mechanisms is not constant, but depends on the carrier lifetime, sample thickness, i.e., the ratio between the sample thickness and the carrier diffusion length, as well as on the modulation frequency of the excitation laser radiation.

For a carrier lifetime of $\tau_1 = 10^{-4}\,\mathrm{s}$, the diffusion length is approximately $L_1$=346.41 $\mu\mathrm{m}$, which means that all analyzed samples can be considered plasma-thin. In this case, photogenerated carriers can diffuse through practically the entire sample volume before recombination, resulting in a relatively homogeneous distribution of recombination heat sources. Consequently, the relative contributions of thermalization and bulk recombination change gradually with increasing frequency, with thermalization becoming dominant at higher frequencies, while the contribution of bulk recombination decreases.

In contrast, for a carrier lifetime of $\tau_2 = 10^{-6}\,\mathrm{s}$, the diffusion length is approximately $L_1$=34.6 $\mu\mathrm{m}$, so that only the 10 and 20 $\mu\mathrm{m}$ samples can be considered plasma-thin, whereas the 50, 100, 200, and 300 $\mu\mathrm{m}$ samples behave as plasma-thick samples. In plasma-thick samples, recombination becomes spatially localized closer to the illuminated surface, making the contribution of recombination heat sources significantly more dependent on frequency. This behavior leads to more pronounced differences between plasma-thin and plasma-thick samples, as well as to the appearance of local maxima and minima in the contributions of thermalization and bulk recombination in the high-frequency region.

A further physical interpretation follows from the dependence of the recombination contribution on the carrier lifetime. Intuitively, the rate of recombination, and consequently the rate of energy release associated with recombination, is approximately related to $R \sim n/\tau$, where n denotes the carrier concentration and $\tau$ the carrier lifetime. Thus, a shorter carrier lifetime corresponds to faster recombination and, consequently, to a larger contribution of recombination heat. Conversely, for a longer carrier lifetime, recombination occurs more slowly, resulting in a reduced recombination contribution and a relatively increased contribution of thermalization.

However, this lifetime dependence should be distinguished from the influence of carrier lifetime on the photothermal signal within a given frequency range. For a fixed carrier lifetime, the relative contribution of recombination can still vary significantly with modulation frequency because the photothermal response depends on the ratio between the characteristic time of the carrier process and the modulation period, expressed by the dimensionless parameter $\omega\tau = 2\pi f \tau$ This frequency dependence is also reflected in the complex diffusion length $L$ , whose frequency-dependent real and imaginary components describe the spatial attenuation and phase delay associated with carrier transport. At low frequencies ($\omega\tau \ll 1$), the system has sufficient time for photogenerated carriers to respond to the modulation and undergo recombination, and the recombination contribution is therefore more pronounced. At high frequencies ($\omega\tau \gg 1$), the rapid modulation limits the ability of the carrier population to follow the excitation, favoring processes that respond more directly to the optical modulation, such as thermalization.

This interpretation explains the characteristic trends observed in the calculated responses. Shorter carrier lifetimes result in a broader frequency range in which recombination remains significant, whereas for longer carrier lifetimes the transition toward thermalization occurs at lower frequencies or the relative recombination contribution becomes weaker. In thicker samples, however, the recombination contribution can become enhanced again at high frequencies due to the spatial distribution of photogenerated carriers and the finite diffusion time associated with carrier transport through the sample.

Particularly significant is the increase in the differences between the illuminated and non-illuminated sides of the sample with increasing frequency. On the illuminated side, the dominant contribution at higher frequencies originates from direct thermalization, since heat is generated in the immediate vicinity of the optical absorption region. On the non-illuminated side, the role of heat transport and carrier diffusion becomes more pronounced, so that the distribution of contributions depends on the ability of carriers to reach the non-illuminated side before recombination. This difference is particularly pronounced in plasma-thick samples, for which the diffusion length represents a limiting factor for the transport of photogenerated carriers.

The analysis of amplitude ratios and phase differences shows that these two characteristics provide complementary information about the photothermal response. While the amplitude characteristics primarily describe the relative intensity of the individual heat sources, the phase characteristics are considerably more sensitive to the dynamics of energy transfer and to the delay between carrier generation, recombination, and heating of the crystal lattice. Consequently, changes in the phase characteristics become more pronounced in frequency regions where a transition occurs between the dominance of recombination processes and direct thermalization.

The most important result of this work is the demonstration that the contribution of bulk recombination to heat generation cannot be neglected even when the surface recombination rate is equal to zero. Although surface recombination is absent under these conditions, photogenerated minority carriers remain within the semiconductor for their finite lifetime and subsequently recombine in the bulk, releasing energy as heat. This finding demonstrates that the commonly adopted assumption that carrier-related contributions vanish for surface recombination rate is not physically justified.

The obtained results indicate that analysis of the frequency response enables the identification of characteristic frequency ranges in which different heat-generation mechanisms dominate. This opens the possibility of using photothermal and photoacoustic methods, together with an appropriate theoretical model, to determine parameters such as minority-carrier lifetime, diffusion length, and surface recombination rate. In particular, the identification of frequency ranges dominated by recombination or thermalization provides a basis for selecting appropriate experimental conditions for the determination of specific carrier-related parameters. Such an approach represents a significant contribution to the non-destructive electrical characterization of semiconductor materials and devices.

## 5. Conclusion

A theoretical model of the photothermal response of optically opaque semiconductors under harmonically modulated optical excitation has been developed by simultaneously accounting for two distinct energy-conversion pathways: direct thermalization of the absorbed optical energy and heat generation through the recombination of photogenerated minority carriers. The model couples carrier diffusion and recombination with thermal diffusion, enabling a systematic analysis of the influence of carrier lifetime, diffusion length, sample thickness, observation side, and modulation frequency on the amplitude and phase characteristics of the photothermal response.

The obtained results demonstrate that the relative contributions of thermalization and bulk recombination are strongly frequency dependent and are governed by the relationship between the carrier diffusion length and the sample thickness. When the diffusion length exceeds the sample thickness, photogenerated carriers explore a large portion of the sample volume before recombination, leading to a gradual transition between recombination-dominated and thermalization-dominated regimes. In contrast, when the diffusion length becomes comparable to or smaller than the sample thickness, pronounced differences emerge between plasma-thin and plasma-thick samples, accompanied by a significantly stronger dependence of the photothermal response on frequency and observation geometry.

Particular attention was devoted to the comparison of the illuminated and non-illuminated sides of the sample. It was shown that the differences between these responses originate from the finite transport lengths of both heat and photogenerated carriers. These differences become increasingly pronounced in plasma-thick samples and therefore contain additional information about carrier transport and recombination processes. Furthermore, the analysis revealed that amplitude and phase characteristics provide complementary information: amplitude ratios primarily reflect the relative intensity of individual heat-generation mechanisms, whereas phase characteristics are significantly more sensitive to the characteristic time scales governing carrier transport, recombination, and energy transfer to the lattice.

One of the most important findings of this work is that the contribution of photogenerated carriers cannot be neglected even when the surface recombination velocity is assumed to be zero. Although surface recombination is absent under these conditions, minority carriers remain present in the semiconductor volume during their finite lifetime and continue to generate heat through bulk recombination. This result demonstrates that the frequently adopted assumption that carrier-related contributions vanish for surface recombination rate equal zero is not physically justified.

From a practical perspective, the results establish guidelines for the design of photothermal and photoacoustic experiments aimed at electrical characterization of semiconductors. Low-frequency measurements and plasma-thin samples maximize sensitivity to carrier recombination and lifetime effects, whereas high-frequency measurements emphasize thermalization processes. The comparison of illuminated and non-illuminated responses, together with simultaneous amplitude and phase analysis, provides a framework for separating thermal and carrier-related contributions and for determining transport parameters such as minority-carrier lifetime, diffusion length, and recombination characteristics. Therefore, the presented approach provides a theoretical basis for the development of non-destructive photothermal methodologies for semiconductor characterization.

Table 2. Recommended measurement conditions for the characterization of carrier-related and thermal processes in semiconductors.

| Measurement objective | Optimal choice | Reason |
|---|---|---|
| Determination of carrier lifetime | Lower frequencies | The recombination process can follow the modulation, and its contribution is therefore maximal |
| Determination of recombination velocity | Thin samples | The influence of diffusion and transport through the bulk is reduced; the signal is more directly related to recombination |
| Determination of carrier diffusion or depth profile | Thick samples | Greater influence of the spatial distribution of carriers and their transit time through the sample |
| Detection of thermalization | Higher frequencies | Recombination becomes weaker, while the direct energy relaxation remains observable |
| Separation of thermalization and recombination | Phase analysis | The phase contains information about the characteristic time constants of the processes |
| Maximum sensitivity to carriers | Illuminated side with low and medium frequencies | This is where the contribution of processes associated with the generated carriers is greatest |

Therefore, the presented approach provides a theoretical basis for the development of non-destructive photothermal methodologies for semiconductor characterization.

Acknowledgment: This work was supported by the Ministry of Science, Technological Development, and Innovation of the Republic of Serbia through the Project contract No 451-0333/2026-03/200017.